\magnification \magstep1
\raggedbottom
\openup 2\jot
\voffset6truemm
\headline={\ifnum\pageno=1\hfill \else
\hfill{\it Local Boundary Conditions in Quantum Supergravity}
\hfill \fi}
\centerline {\bf LOCAL BOUNDARY CONDITIONS IN}
\centerline {\bf QUANTUM SUPERGRAVITY}
\vskip 1cm
\leftline {\it Giampiero Esposito}
\vskip 1cm
\leftline {Istituto Nazionale di Fisica Nucleare, Sezione di Napoli}
\leftline {Mostra d'Oltremare Padiglione 20, 80125 Napoli, Italy;}
\leftline {Dipartimento di Scienze Fisiche}
\leftline {Mostra d'Oltremare Padiglione 19, 80125 Napoli, Italy.}
\vskip 1cm
\noindent
{\bf Abstract.} When quantum supergravity is studied on
manifolds with boundary, one may consider local boundary conditions
which fix on the initial surface the whole primed part
of tangential components of gravitino perturbations, and fix
on the final surface the whole unprimed part of tangential
components of gravitino perturbations. This paper studies such
local boundary conditions in a flat Euclidean background 
bounded by two concentric 3-spheres. It is shown that, 
as far as transverse-traceless perturbations are concerned,
the resulting contribution to $\zeta(0)$ vanishes
when such boundary data are set to zero,
exactly as in the case when non-local boundary conditions of the
spectral type are imposed. These properties may be used to
show that one-loop finiteness of massless supergravity models
is only achieved when two boundary 3-surfaces occur, and there 
is no exact cancellation of the contributions of gauge and
ghost modes in the Faddeev-Popov path integral. In these
particular cases, which rely on the use of covariant
gauge-averaging functionals, pure gravity is one-loop 
finite as well.
\vskip 100cm
The problem of a consistent formulation of quantum supergravity
on manifolds with boundary is still receiving careful
consideration in the current literature [1--4]. In particular,
many efforts have been produced to understand whether simple
supergravity is one-loop finite (or even finite to all orders
of perturbation theory [3]) in the presence of boundaries.
In the analysis of such an issue, the first problem consists,
of course, in a careful choice of boundary conditions. For
massless gravitino potentials, which are the object of our
investigation, these may be non-local of the spectral type
[5] or local [1--4].

In the former case the idea is to fix at the boundary half of
the gravitino potential. On the final surface $\Sigma_{F}$ one
can fix those perturbative modes which multiply harmonics 
having positive eigenvalues of the intrinsic three-dimensional
Dirac operator $\cal D$ of the boundary. On the initial surface
one can instead fix those gravitino modes which multiply
harmonics having negative eigenvalues of the intrinsic
three-dimensional Dirac operator of the boundary. What is
non-local in this procedure is the separation of the spectrum
of a first-order elliptic operator (our $\cal D$) into a positive
and a negative part. This leads to a sort of positive- and
negative-frequency split which is typical for scattering
problems [3], but may also be applied to the analysis of 
quantum amplitudes in finite regions [4]. 

Our paper deals instead with the latter choice, i.e. local
boundary conditions for quantum supergravity. By this one 
usually means a formulation where complementary projection
operators act on gravitational and spin-${3\over 2}$
perturbations. Local boundary conditions of this type were
investigated in Refs. [6,7], and then applied to quantum
cosmological backgrounds in Refs. [1,4,8--10].

More recently, another choice of local boundary conditions 
for gravitino perturbations has been considered in Ref. 
[3]. Using two-component spinor notation, and referring 
the reader to Refs. [3,11] for notation and background material,
we here represent the spin-${3\over 2}$ potential by a pair
of independent spinor-valued one-forms
$\Bigr(\psi_{\mu}^{A},{\widetilde \psi}_{\mu}^{A'}\Bigr)$ in
a Riemannian 4-manifold which is taken to be flat
Euclidean 4-space bounded by two concentric 3-spheres
[2,4]. Denoting by $S_{I}$ and $S_{F}$ the boundary 3-spheres,
with radii $a$ and $b$ respectively (here $b>a$), the local
boundary conditions proposed in Ref. [3] read in our case
($i=1,2,3$)
$$
\Bigr[{\widetilde \psi}_{i}^{A'}\Bigr]_{S_{I}}
=F_{i}^{A'} \; ,
\eqno (1)
$$
$$
\Bigr[\psi_{i}^{A}\Bigr]_{S_{F}}=H_{i}^{A} \; ,
\eqno (2)
$$
where $F_{i}^{A'}$ and $H_{i}^{A}$ are boundary
data which may or may not satisfy the classical constraint
equations [3]. With the choice (1) and (2), the whole primed
part of the tangential components of the spin-${3\over 2}$
potential is fixed on $S_{I}$, and the whole unprimed part
of the tangential components of the spin-${3\over 2}$ potential
is fixed on $S_{F}$.

In a Hamiltonian analysis, $\psi_{0}^{A}$ and 
${\widetilde \psi}_{0}^{A'}$ are Lagrange multipliers, and hence
boundary conditions for them look un-natural in a one-loop
calculation [3], especially if one is interested in reduction
to transverse-traceless (TT) gravitino modes (usually regarded
as the {\it physical} part of gravitinos). In a covariant
path-integral analysis, however, one cannot disregard the issue
of boundary conditions on normal components of gravitinos.
We shall thus post-pone the discussion of this point, and we
will focus on the TT sector of the boundary conditions (1)
and (2).

With the notation of Ref. [1], the expansion in harmonics on
3-spheres of the TT part of tangential components of
spin-${3\over 2}$ perturbations reads
$$
\psi_{i}^{A}={\tau^{-{3\over 2}}\over 2\pi}\sum_{n=0}^{\infty}
\sum_{p,q=1}^{(n+1)(n+4)}\alpha_{n}^{pq}
\Bigr[m_{np}(\tau)\beta^{nqACC'}
+{\widetilde r}_{np}(\tau){\overline \mu}^{nqACC'}\Bigr]
e_{CC'i} \; ,
\eqno (3)
$$
$$
{\widetilde \psi}_{i}^{A'}={\tau^{-{3\over 2}}\over 2\pi}
\sum_{n=0}^{\infty}\sum_{p,q=1}^{(n+1)(n+4)}\alpha_{n}^{pq}
\Bigr[{\widetilde m}_{np}(\tau)
{\overline \beta}^{nqA'C'C}+r_{np}(\tau)
\mu^{nqA'C'C}\Bigr]e_{CC'i} \; ,
\eqno (4)
$$
where $\beta^{nqACC'} \equiv -\rho^{nq(ACD)}n_{D}^{\; \; C'}$
and $\mu^{nqA'C'C} \equiv -\sigma^{nq(A'C'D')}
n_{\; \; D'}^{C}$. Of course, round brackets denote complete
symmetrization over spinor indices, and $n_{\; \; D'}^{C}$ is
obtained from the Euclidean normal as $n_{\; \; D'}^{C}
=i {_{e}n_{\; \; D'}^{C}}$. Variation of the TT gravitino action
yields, for all integer $n \geq 0$, the following eigenvalue 
equations for gravitino modes:
$$
\biggr({d\over d\tau}-{(n+5/2)\over \tau}\biggr)x_{np}
=E_{np} \; {\widetilde x}_{np} \; ,
\eqno (5)
$$
$$
\biggr(-{d\over d\tau}-{(n+5/2)\over \tau}\biggr)
{\widetilde x}_{np}=E_{np} \; x_{np} \; ,
\eqno (6)
$$
where $x_{np}=m_{np}$ and ${\widetilde x}_{np}=
{\widetilde m}_{np}$, or $x_{np}=r_{np}$ and
${\widetilde x}_{np}={\widetilde r}_{np}$. The Eqs. (5) and
(6) lead to the following basis functions in terms of
modified Bessel functions (hereafter $M \equiv E_{np}$ for
simplicity of notation, while $\beta_{1,n}$ and $\beta_{2,n}$
are some constants):
$$
m_{np}(\tau)=\beta_{1,n}\sqrt{\tau}I_{n+2}(M\tau)
+\beta_{2,n}\sqrt{\tau}K_{n+2}(M\tau) \; ,
\eqno (7)
$$
$$
{\widetilde m}_{np}(\tau)=\beta_{1,n}\sqrt{\tau}I_{n+3}(M\tau)
-\beta_{2,n}\sqrt{\tau}K_{n+3}(M\tau) \; ,
\eqno (8)
$$
$$
r_{np}(\tau)=\beta_{1,n}\sqrt{\tau}I_{n+2}(M\tau)
+\beta_{2,n}\sqrt{\tau}K_{n+2}(M\tau) \; ,
\eqno (9)
$$
$$
{\widetilde r}_{np}(\tau)=\beta_{1,n}\sqrt{\tau}I_{n+3}(M\tau)
-\beta_{2,n}\sqrt{\tau}K_{n+3}(M\tau) \; .
\eqno (10)
$$
By virtue of (1) and (2), these modes obey the boundary
conditions
$$
{\widetilde m}_{np}(a)=A_{n} \; ,
\eqno (11a)
$$
$$
r_{np}(a)=A_{n} \; ,
\eqno (11b)
$$
$$
m_{np}(b)=B_{n} \; ,
\eqno (12a)
$$
$$
{\widetilde r}_{np}(b)=B_{n} \; ,
\eqno (12b)
$$
where $A_{n}$ and $B_{n}$ are constants resulting from the
boundary data $F_{i}^{A'}$ and $H_{i}^{A}$ respectively. The
boundary conditions (11) and (12) lead therefore, 
if $A_{n}=B_{n}=0$, $\forall n$, to the eigenvalue condition
$$
I_{n+3}(Mr)K_{n+2}(Mr)+I_{n+2}(Mr)K_{n+3}(Mr)=0 \; ,
\eqno (13)
$$
where $r=a$ or $b$. 

We can now apply $\zeta$-function regularization to evaluate 
the resulting TT contribution to the one-loop divergence,
following the algorithm developed in Ref. [12]. The basic steps
are as follows. Let us denote by $f_{l}$ the function occurring
in the equation obeyed by the eigenvalues by virtue of boundary
conditions, after taking out fake roots (e.g. $x=0$ is a fake
root of order $n$ of the Bessel function $I_{n}$). Let $d(l)$ 
be the degeneracy of the eigenvalues parametrized by the integer
$l$. One can then define the function
$$
I(M^{2},s) \equiv \sum_{l=l_{0}}^{\infty} d(l) l^{-2s}
\log f_{l}(M^{2}) \; .
\eqno (14)
$$
Such a function admits an analytic continuation to the
complex-$s$ plane as a meromorphic function with a simple pole
at $s=0$, in the form
$$
``I(M^{2},s)"={I_{\rm pole}(M^{2})\over s}
+I^{R}(M^{2})+{\rm O}(s) \; .
\eqno (15)
$$
The function $I_{\rm pole}(M^{2})$ is the residue at $s=0$, 
and makes it possible to obtain the $\zeta(0)$ value as
$$
\zeta(0)=I_{\rm log}+I_{\rm pole}(M^{2}=\infty)
-I_{\rm pole}(M^{2}=0) \; ,
\eqno (16)
$$
where $I_{\rm log}$ is the coefficient of the $\log M$ term in
$I^{R}$ as $M \rightarrow \infty$. Moreover, $I_{\rm pole}(\infty)$
coincides with the coefficient of ${1\over l}$ in the expansion 
as $l \rightarrow \infty$ of ${1\over 2}d(l) \log[\rho_{\infty}(l)]$,
where $\rho_{\infty}(l)$ is the $l$-dependent term in the eigenvalue
condition as $M \rightarrow \infty$ and $l \rightarrow \infty$. 
The $I_{\rm pole}(0)$ value is instead obtained as the coefficient
of ${1\over l}$ in the expansion as $l \rightarrow \infty$ of
${1\over 2}d(l) \log[\rho_{0}(l)]$, where $\rho_{0}(l)$ is the
$l$-dependent term in the eigenvalue condition as 
$M \rightarrow 0$ and $l \rightarrow \infty$.

In our problem, using the limiting form of Bessel functions when
the argument tends to zero, one finds that the left-hand side 
of (13) is proportional to $M^{-1}$ as $M \rightarrow 0$. Hence
one has to multiply by $M$ to get rid of fake roots. Moreover,
in the uniform asymptotic expansion of Bessel functions as
$M \rightarrow \infty$ and $n \rightarrow \infty$, both $I$
and $K$ functions contribute a ${1\over \sqrt{M}}$ factor. 
These properties imply that $I_{\rm log}$ vanishes
(hereafter $l \equiv n+1$):
$$
I_{\rm log}={1\over 2}\sum_{l=1}^{\infty}2l(l+3)
\Bigr(1-1/2-1/2 \Bigr)=0 \; .
\eqno (17)
$$
Moreover, $I_{\rm pole}(\infty)$ vanishes since there is no
$l$-dependent coefficient in the uniform asymptotic expansion of
(13) as $M \rightarrow \infty$ and $l \rightarrow \infty$. 
Last, $I_{\rm pole}(0)$ vanishes as well, since the limiting
form of (13) as $M \rightarrow 0$ and $l \rightarrow \infty$
is ${1\over r}M^{-1}$. One thus finds for gravitinos
$$
\zeta_{TT}(0)=0 \; .
\eqno (18)
$$
 
It is now clear that local boundary conditions for 
$\psi_{0}^{A}$ and ${\widetilde \psi}_{0}^{A'}$ along the
same lines of (1) and (2), i.e. (here $\kappa^{A'}$ and
$\mu^{A}$ are some boundary data)
$$
\Bigr[{\widetilde \psi}_{0}^{A'}\Bigr]_{S_{I}}
=\kappa^{A'} \; ,
\eqno (19)
$$
$$
\Bigr[\psi_{0}^{A}\Bigr]_{S_{F}}=\mu^{A} \; ,
\eqno (20)
$$
give again a vanishing contribution to $\zeta(0)$ in this
two-boundary problem if $\kappa^{A'}$ and $\mu^{A}$
are set to zero, since the resulting eigenvalue condition
is analogous to (13) with $n$ replaced by $n-1$. 

A suitable set of local boundary conditions on metric perturbations
$h_{\mu \nu}$ are the ones considered by Luckock, Moss and 
Poletti [6--8]. In our problem, denoting by $g_{\mu \nu}$ the
background 4-metric, they read [9,10]
$$
\Bigr[h_{ij}\Bigr]_{\partial M}=0 \; ,
\eqno (21)
$$
$$
\biggr[{\partial h_{00}\over \partial \tau}
+{6\over \tau}h_{00}-{\partial \over \partial \tau}
\Bigr(g^{ij}h_{ij}\Bigr)\biggr]_{\partial M}=0 \; ,
\eqno (22)
$$
$$
\Bigr[h_{0i}\Bigr]_{\partial M}=0 \; ,
\eqno (23)
$$
while the ghost 1-form is subject to the mixed boundary
conditions [10]
$$
\Bigr[\varphi_{0}\Bigr]_{\partial M}=0 \; ,
\eqno (24)
$$
$$
\biggr[{\partial \varphi_{i} \over \partial \tau}
-{2\over \tau}\varphi_{i} \biggr]_{\partial M}=0 \; .
\eqno (25)
$$
As shown in Ref. [9], graviton TT modes contribute
$$
\zeta_{TT}(0)=-5 \; ,
\eqno (26)
$$
while, using a covariant gauge-averaging functional of the
de Donder type, gauge and ghost modes contribute [9]
$$
\zeta(0)_{\rm {gauge \; and \; ghost}}=5 \; .
\eqno (27)
$$
One thus finds that the one-loop path integral for the
gravitational sector, including TT, gauge and ghost modes,
gives a vanishing contribution to the one-loop divergence.

What is left are gauge and ghost modes for gravitino
perturbations. In general, their separate values depend on
the gauge-averaging functional being used (only the full 
one-loop divergence should be gauge-independent [13]).
However, on general ground, since in our flat Euclidean
background all possible contributions to $\zeta(0)$ involve 
surface integrals of terms like [3,14]
$$
{\rm Tr}(K^{3}) \; , \; ({\rm Tr}K)({\rm Tr}K^{2})
\; , \; ({\rm Tr}K)^{3} \; ,
$$
$K$ denoting the extrinsic-curvature tensor of the boundary, 
any dependence on the 3-spheres radii $a$ and $b$ 
disappears after integration over $\partial M$. Moreover,
{\it if} the formalism is gauge-independent ax expected, the
one-loop result can only coincide with the one found in Sec. V 
of Ref. [4] in the axial gauge:
$$
\zeta_{3\over 2}(0)=0 \; .
\eqno (28)
$$
The following concluding remarks are now in order:
\vskip 0.3cm
\noindent
(i) The one-loop finiteness suggested by our analysis
does not seem to contradict the results of Ref. [4]. What
is shown in Ref. [4] is instead that, when flat Euclidean 
4-space is bounded by {\it only one} 3-sphere, simple
supergravity fails to be one-loop finite (either spectral
boundary conditions with non-covariant gauge, or local boundary
conditions in covariant gauge). This is {\it not} our background.
Moreover, the two-boundary problem of Ref. [4] was studied in
a non-covariant gauge of the axial type, when the Faddeev-Popov
path integral is supplemented by the integrability condition
for the eigenvalue equations on graviton and gravitino
perturbations. This quantization was found to pick out TT
modes only [4], but differs from the scheme proposed in
our paper.
\vskip 0.3cm
\noindent
(ii) The result (28) is crucial for one-loop finiteness to
hold (when combined with (26) and (27)), 
and its explicit proof has not yet been achieved.
\vskip 0.3cm
\noindent
(iii) As a further check of one-loop finiteness (or of its
lack), one should now perform two-boundary calculations of
$\zeta(0)$ when Luckock-Moss-Poletti local boundary conditions
are imposed on gravitino perturbations.
\vskip 0.3cm
\noindent
(iv) Yet another check might be obtained by combining 
Barvinsky boundary conditions for pure gravity [10,15] in the
de Donder gauge (these are completely invariant under infinitesimal
diffeomorphisms) with local or non-local boundary conditions 
for gravitinos in covariant gauges [13].
\vskip 0.3cm
\noindent
(v) The calculations performed in Refs. [4,9,10,16] and in
our paper show that, when cancellation of the effects of gauge
and ghost modes is achieved, only the effects of TT modes 
survive, and hence both pure gravity and simple supergravity
are not even one-loop finite. By contrast, if the effects of
gauge and ghost modes do not cancel each other exactly (e.g.
by using covariant gauge-averaging functionals in the 
Faddeev-Popov path integral), then both pure gravity and 
simple supergravity may turn out to be one-loop finite in
the presence of two bounding 3-spheres.
\vskip 0.3cm
\noindent
(vi) A deep problem is the relation between the
Hamiltonian analysis of Ref. [3], where auxiliary fields
play an important role but ghost fields are not studied,
and the path-integral approach of Ref. [4], where ghost
fields are analyzed in detail but auxiliary fields are not
found to affect the one-loop calculation.

All this adds evidence in favour of quantum cosmology being
able to lead to new perspectives in Euclidean quantum gravity
and quantum supergravity.
\vskip 1cm
I am much indebted to A. Yu. Kamenshchik
for scientific collaboration on $\zeta$-function regularization
and boundary conditions in one-loop quantum cosmology. I am
also grateful to P. D'Eath for correspondence and conversations
on quantum supergravity.
\vskip 1cm
\leftline {\bf References}
\vskip 1cm
\item {[1]}
G. Esposito, Quantum gravity, quantum cosmology and Lorentzian
geometries, Lecture Notes in Physics, new series m: Monographs,
vol. m12, second corrected and enlarged edition (Springer-Verlag,
Berlin, 1994).
\item {[2]}
G. Esposito, G. Gionti, A.Yu. Kamenshchik, I.V. Mishakov and
G. Pollifrone, Int. J. Mod. Phys. D 4 (1995) 735.
\item {[3]}
P. D. D'Eath, Supersymmetric quantum cosmology (Cambridge
University Press, Cambridge, 1996).
\item {[4]}
G. Esposito and A.Yu. Kamenshchik, Phys. Rev. D 54 (1996)
(hep-th 9604182).
\item {[5]}
P.D. D'Eath and G. Esposito, Phys. Rev. D 44 (1991) 1713.
\item {[6]}
H.C. Luckock and I.G. Moss, Class. Quantum Grav. 6 (1989) 1993.
\item {[7]}
H.C. Luckock, J. Math. Phys. 32 (1991) 1755.
\item {[8]}
I.G. Moss and S. Poletti, Nucl. Phys. B 341 (1990) 155.
\item {[9]}
G. Esposito, A.Yu. Kamenshchik, I.V. Mishakov and G. Pollifrone,
Phys. Rev. D 50 (1994) 6329.
\item {[10]}
G. Esposito, A.Yu. Kamenshchik, I.V. Mishakov and G. Pollifrone,
Phys. Rev. D 52 (1995) 3457.
\item {[11]}
P.D. D'Eath, Phys. Rev. D 29 (1984) 2199.
\item {[12]}
A.O. Barvinsky, A.Yu. Kamenshchik and I.P. Karmazin,
Ann. Phys. (N.Y.) 219 (1992) 201.
\item {[13]}
R. Endo, Class. Quantum Grav. 12 (1995) 1157.
\item {[14]}
P.B. Gilkey, Invariance theory, the heat equation, and the
Atiyah-Singer index theorem (Chemical Rubber Company,
Boca Raton, 1995).
\item {[15]}
A.O. Barvinsky, Phys. Lett. B 195 (1987) 344.
\item {[16]}
I.G. Avramidi, G. Esposito and A.Yu. Kamenshchik, Class.
Quantum Grav. 13 (1996) 2361.

\bye